\documentclass[aps,preprint,showpacs]{revtex4}
\usepackage{graphicx}
\usepackage{bm}
\usepackage{amsmath}

\begin{document}

\title{The Orbital Selective Mott Transition in a Three Band Hubbard model:
a Slave Boson Mean Field Study}
\author{Xi Dai$^{a,b}$, Gabriel Kotliar$^c$, Zhong Fang$^a$}
\affiliation{$^a$ Institute of Physics, Chinese Academy of Sciences, Beijing, China}
\affiliation{$^b$Department of Physics, University of Hong Kong,Hong Kong, China}
\email[E-mail:]{daix@hkucc.hku.hk}
\affiliation{$^c$Department of Physics and Center for Material Theory, Rutgers
University, Piscataway, NJ 08854, USA}
\date{\today }

\begin{abstract}
In the present paper, we systematically studied the possible Orbital
selective Mott transition (OSMT) in the $t_{2g}$ system, which has three
orbital degeneracy. The slave Boson mean field theory is generalized to the
three band systems with full Hund's rule coupling including spin flip and
pair hopping terms. A new type of Mott transition which is driven by the
crystal field splitting or the lattice distortion is found is this model. We
argue that this new type of Mott transition may relate to the puzzling phase
in $Sr_{x}Ca_{2-x}RuO_4$.
\end{abstract}

\maketitle

\section{\protect\bigskip Introduction}

The charge, spin, orbital and lattice are the main ingredient of the physics
in transition metal oxides\cite{orbital_review}. The interplay and
competition among them are the key points to understand the interesting
physical properties in transition metal oxides. On the other hand, Mott
transition plays an important role in understanding the correlation effect
in the solid state\cite{Mott}. One of the important debating issues in the
study of Mott transition is wether the Mott transition in realistic material
is purely interaction driven or the orbital and lattice degrees of freedom
also play some role\cite{VO2,Millis}. Although most of the theoretical
studies on the Mott transition are based on the single band Hubbard model,
most of the realistic materials which undergo Mott transition have more than
one active orbit\cite{Mott,Anisimov}. The recent studies show that the Mott
transition in the multi-orbital system is not the simple generation of its
single band version, instead it has some interesting features which are
unique in the multi-orbital case\cite{Capone,multi}. It is even more
complicated if the lattice distortion get involved and play some role in the
Mott transition. Therefore it is very interesting to study how the orbital
fluctuation and lattice distortion will affect the nature of Mott
transition. Is there any new physics near the Mott transition point induced
by the interplay between the charge, spin, orbital and lattice?

The orbital selective Mott transition (OSMT) is a very important feature for
the Mott transition in the multi-orbital system and is first proposed by
Anisimov in the study of $Sr_{x}Ca_{2-x}RuO_4$.\cite{Anisimov,exp1,exp2,exp3}
The orbital degree of freedom is involved in the OSMT and the Mott
transition in different orbital happens individually because of the lift of
the orbital degeneracy. In crystal, basically there are two origins which
can induce the asymmetry in the orbital space. The first one is the breaking
of the local rotational symmetry in the crystal, which makes the different
orbits pointing along different direction have different band width. The
second one is the crystal field splitting generated by the lattice
distortion, which removes the degeneracy of the energy levels. A two-band
generalized Hubbard model with half filling has been proposed\cite%
{Koga,Liebsch} to study the basic features of OSMT. Most of the studies on
this two-band model using the dynamical mean field theory (DMFT)\cite{DMFT}
reach the following common conclusions\cite{Koga,Liebsch,Georges}. i)The
OSMT is induced by the band width asymmetry of the different orbits. The
crystal field splitting will reduce rather than enhance the tendency to
OSMT. ii)The orbital selective Mott Phase (OSMP), in which one orbit is
already localized while the other one keeps itinerary, will be greatly
enhanced when the symmetry of the local interaction is lowered by including
the Hund's rule coupling terms.

Although the two-band model has been extensively studied in recent years, it
may not be directly applied to the situation of $Sr_{x}Ca_{2-x}RuO_4$,
because the effective model here is the three-band model with four total
electrons\cite{Anisimov,Fang}. The most interesting feature of the OSMT in $%
Sr_{x}Ca_{2-x}RuO_4$ is that the OSMT will be accompanied by the
change of orbital polarization. Before the OSMT, as predicted by
LDA calculation and conformed by ARPES \cite{Fang,exp3}, the four
electrons are almost evenly distributed among the three $t_{2g}$
orbits. Therefore the average occupation number of each orbital is
$4/3$ before the OSMT. In order to have OSMP, $1/3$ of the
electron has to be moved from the localized orbital to the
non-localized ones. Thus the OSMT is always accompanied by the
charge redistribution or the change of orbital polarization. Since
the crystal field splitting strongly couples to the orbital
polarization, it will play a very important role in the OSMT for
the three-band model.

From the methodology point of view, it is very difficult to apply
the DMFT to the three band Hubbard model, because of the
limitation of the computation ability. In the present paper, we
generalize the slave Boson mean field theory\cite{gabi,gabi2}
which is very successful in the single band Hubbard model to the
three band model with both local repulsive interaction and Hund's
rule coupling. The slave boson method is closely related to the
Gutzwiller approximation,  which was recently extended to
incorporate the effects of  multiplets\cite{GW}. We give the
detail description of the slave Boson mean field theory in section
II and benchmark it in the two band model with other approaches in
section III. Finally we discuss the complete phase diagram for the
three band Hubbard model with Hund's rule coupling in section IV
and make some important conclusions in section V.

\section{The Hamiltonian and the slave boson mean field approximation}

The general Hamiltonian describing the $t_{2g}$ system can be written as

\begin{eqnarray}
H_{Total} &=&H_{0}+H_{I}  \notag \\
H_{0} &=&\sum_{k,\alpha \sigma }\epsilon _{k,\alpha }C_{k\alpha \sigma
}^{\dag }C_{k\alpha \sigma }+\sum_{k,\alpha \sigma }\Delta _{\alpha
}C_{k\alpha \sigma }^{\dag }C_{k\alpha \sigma }  \notag \\
H_{I} &=&U\sum_{i,\alpha }n_{i\alpha \uparrow }n_{i\alpha \downarrow
}+U^{\prime }\sum_{i,\sigma \sigma ^{\prime },\alpha <\beta }n_{i\alpha
\sigma }n_{i\beta \sigma ^{\prime }}-J_{z}\sum_{i,\alpha <\beta ,\sigma
}n_{i\alpha \sigma }n_{i\beta \sigma }  \notag \\
&&-J_{xy}\sum_{i,\alpha <\beta }\left[ C_{i\alpha \uparrow }^{\dag
}C_{i\alpha \downarrow }C_{i\beta \downarrow }^{\dag }C_{i\beta \uparrow
}+C_{i\alpha \downarrow }^{\dag }C_{i\alpha \uparrow }C_{i\beta \uparrow
}^{\dag }C_{i\beta \downarrow }\right]  \notag \\
&&-J_{xy}\sum_{i,\alpha <\beta }\left[ C_{i\alpha \uparrow }^{\dag
}C_{i\alpha \downarrow }^{\dag }C_{i\beta \uparrow }C_{i\beta \downarrow
}+C_{i\beta \uparrow }^{\dag }C_{i\beta \downarrow }^{\dag }C_{i\alpha
\uparrow }C_{i\alpha \downarrow }\right]  \label{3band}
\end{eqnarray}

where $\alpha $ and $\sigma $ are the orbital and spin indices respectively.
$U$ describes the on-site Coulomb interaction term between two electrons in
the same orbit but with opposite spin. While $U^{\prime }$ describes the
on-site interaction term for two electrons reside in different orbits. $%
\Delta _{\alpha }$ represent the crystal field splitting of the three $%
t_{2g} $ orbits. We divide the Hund's rule coupling terms into three parts.
The first one is the $J_{z}$ term which describes the longitudinal part of
the Hund's rule coupling which only involves density-density coupling. While
the other two $J_{xy}$ terms describe the spin flip and pair hopping
processes respectively. In the present study, we assume the system has
approximately the cubic symmetry which gives the following constraint $%
U=U^{\prime}+2J$ with $J_{z}=J_{xy}=J$.

Following reference \cite{gabi} and \cite{gabi2}, we first
diagonalize the local Hamiltonian $H_{I}$ by a set of local bases
$\left\vert \mu \right\rangle $ and for each of them we define a
slave boson operator $l_{\mu }^{+}$. Then we can express the
physical electron operator in terms of the pseudo fermions
$f_{i\alpha \sigma }^{+}$ and slave bosons as

\begin{equation}
C_{i\alpha \sigma }^{+}=Z_{i\alpha }^{+}f_{i\alpha \sigma }^{+}
\end{equation}%
,

where

\begin{equation}
Z_{i\alpha \sigma }^{+}=\sum_{\mu \nu }D_{\nu \mu }^{\alpha \sigma \ast
}L_{i\alpha \sigma }l_{i\mu }^{+}l_{i\nu }R_{i\alpha \sigma }
\end{equation}

with

\begin{eqnarray}
L_{\alpha \sigma } &=&\frac{1}{\sqrt{n_{i\alpha \sigma }}} \\
R_{\alpha \sigma } &=&\frac{1}{\sqrt{1-n_{i\alpha \sigma }}}  \notag
\end{eqnarray}

and $D_{\nu \mu }^{\alpha \sigma }=\left\langle \nu \right\vert
C_{\alpha \sigma }\left\vert \mu \right\rangle $ being the matrix
elements of the electron operator represented in terms of the
local atomic states.

Therefore the original Hamiltonian can be written in terms of the slave
particles as

\bigskip

\begin{equation}
H_{sb}=\sum_{ij,\alpha \sigma }t_{ij,\alpha \sigma }f_{i\alpha \sigma
}^{+}f_{j\alpha \sigma }Z_{i\alpha \sigma }^{+}Z_{j\alpha \sigma
}+H.C.+\sum_{i,\alpha \sigma }\Delta _{\alpha }f_{i\alpha \sigma
}^{+}f_{i\alpha \sigma }+\sum_{i,\mu }E_{\mu }l_{i\mu }^{+}l_{i\mu }
\end{equation}

with two following local constraints
\begin{eqnarray}
\sum_{\mu }l_{i\mu }^{+}l_{i\mu } &=&1  \label{const1} \\
\sum_{\mu }\eta _{\alpha \sigma ,\mu }l_{i\mu }^{+}l_{i\mu } &=&f_{i\alpha
\sigma }^{+}f_{i\alpha \sigma }  \label{const2}
\end{eqnarray}

, where $\eta _{\alpha \sigma ,\mu }=\sum_{\nu }D_{\nu \mu
}^{\alpha \sigma \ast }D_{\nu \mu }^{\alpha \sigma }$ is the
beverage particle number with the orbital $\alpha $ and spin
$\sigma $ for the configuration $\mu $.

In the mean field approach, we treat all the boson operators $l_{\mu }$ and $%
l_{\mu }^{+}$ as c-numbers which will be determined by
minimization the ground state energy. The local constraints
\ref{const1} and \ref{const2} will be released to be global ones,
which can be satisfied by two Lagrange multiples $\lambda _{1}$
and $\lambda _{2}$. Therefore we obtain the following mean field
Hamiltonian

\begin{eqnarray}
H_{mf} &=&\sum_{ij,\alpha \sigma }t_{ij,\alpha \sigma }f_{i\alpha \sigma
}^{+}f_{j\alpha \sigma }Z_{\alpha \sigma }^{\ast }Z_{\alpha \sigma
}+H.C.+\sum_{i,\alpha \sigma }\left( \Delta _{\alpha }-\mu _{f}+\lambda
_{1}\right) f_{i\alpha \sigma }^{+}f_{i\alpha \sigma } \\
&&+\sum_{i,\mu }\left( E_{\mu }-\xi _{\mu }\right) l_{\mu }^{+}l_{\mu
}+\sum_{i,\mu }\lambda _{2}\left( l_{\mu }^{+}l_{\mu }-1\right)  \notag
\end{eqnarray}

with $\xi _{\mu }=\sum_{\alpha \sigma }\eta _{\alpha \sigma ,\mu }\lambda
_{1}$. The mean field solution can be determined by the stationary condition
of the ground state energy,

\begin{eqnarray}
\frac{\partial E\left[ l_{\mu },l_{\mu }^{+},\lambda _{1},\lambda _{2}\right]
}{\partial l_{\mu }} &=&0  \notag \\
\frac{\partial E\left[ l_{\mu },l_{\mu }^{+},\lambda _{1},\lambda _{2}\right]
}{\partial \lambda _{i}} &=&0
\end{eqnarray}

For the three band model discussed here, we have to minimize the
mean field ground state energy respect to all the 64 independent
$l_{\mu }$ as well as two Lagrange multipliers and one chemical
potential. Therefore the main difficulty here is how to reach the
saddle point efficiently in the 67 dimensional parameter space.
The simple iterative procedure is very hard to converge and is
almost useless in practice. In the present paper, we solve this
problem by a so called "adiabatic solution searching" procedure. \
First we solve the mean field self consistent equation for a
symmetric 3-band model with equal band width, no crystal field
splitting and $J=0$ ($SU(6)$ symmetric case), which is easy due to
the high degeneracies of the local configurations. Then we tune
the parameters step by step towards the realistic parameters we
are studying. For each step, we use the converged solution of the
previous step as the initial value for the iteration. Since we
only change the parameters slightly at each step, the new self
consistent solution should be very close to the previous one and
the convergence can be reached very soon. Therefore when the
parameters evolute slowly to their actual values, we adiabatically
obtained the solution.

\bigskip

\section{Benchmarks}

\begin{figure}[tbp]
\begin{center}
\includegraphics[width=10cm,angle=0,clip=]{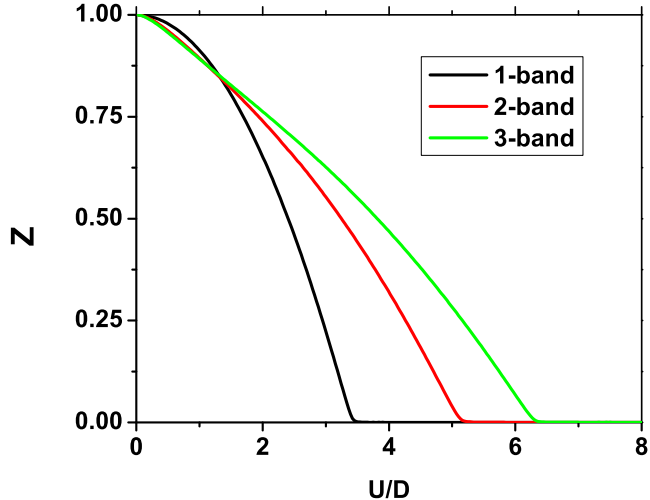}
\end{center}
\caption{The quasi-particle weight as the function of U/D for 1,2, and 3
band Hubbard model. }
\label{fig1}
\end{figure}

In this section, we benchmark our present mean field approach with
other numerical approaches. First we apply the slave Boson mean
field theory to the multi-band Hubbard model with SU(N) symmetry,
where both the crystal field splitting and Hund's rule coupling
has been set to zero. The non-interacting density of states of
each band is chosen to be semi-circle with half bandwidth $D$,
which also applies to all the models studied in this paper. The
quasi-particle weight reduction as the function of $U/D$ is
plotted in figure \ref{fig1}. The Mott transitions can be seen
clearly where the quasi-particle weight vanishes. The critical interaction $%
U_{c}$ for 1,2 and 3 band model are found to be $3.42D$, $5.1D$
and $6.27D$, which is in good agreement with the results obtained
by dynamical mean field theory (DMFT) and other slave particle
mean field theories\cite{Georges}.

\begin{figure}[tbp]
\begin{center}
\includegraphics[width=10cm,angle=0,clip=]{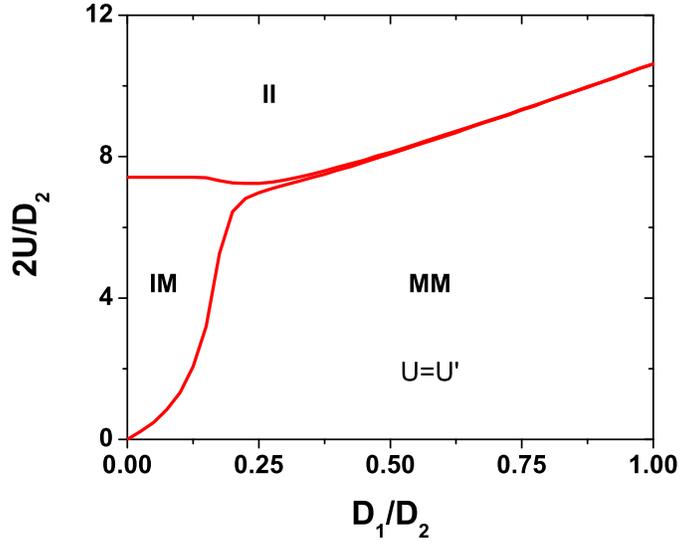}
\end{center}
\caption{The phase diagram for the 2-band Hubbard model without Hund's rule
coupling. }
\label{fig2}
\end{figure}

We then apply the slave boson mean field approach to the 2-band
Hubbard model with unequal bandwidth $D_{2}$ and $D_{1}$ without
crystal field splitting. The total occupation number of the
electrons are fixed to be two, which keeps the particle hole
symmetry for this model. The DMFT study on this model reveals an
interesting new phase called orbital selective Mott phase (OSMP),
where the narrow band is in the Mott insulator phase while the
wide band still remains metallic\cite{Koga,Liebsch}. This
simplified model is first proposed to
explain the possible orbital selective Mott transition in the $%
Ca_{2-x}Sr_{x}RuO_{4}$. Since it has been widely studied by DMFT as well as
mean field approaches, it is quite suitable to be the benchmark of our
approach. The DMFT studies on this model show that the OSMT does not occur
without the Hund's rule coupling term for $D_{1}/D_{2}=0.5$. This is because
the local interaction has the $SU(4)$ symmetry without the Hund's rule
coupling terms, which prevents the OSMT from occurring due to enhancement of
the orbital fluctuation. We plot the phase diagram obtained by the slave
boson mean field theory in figure \ref{fig2} with y-axis being the
interaction strength and x-axis being the radio of two band width.

\begin{figure}[tbp]
\begin{center}
\includegraphics[width=10cm,angle=0,clip=]{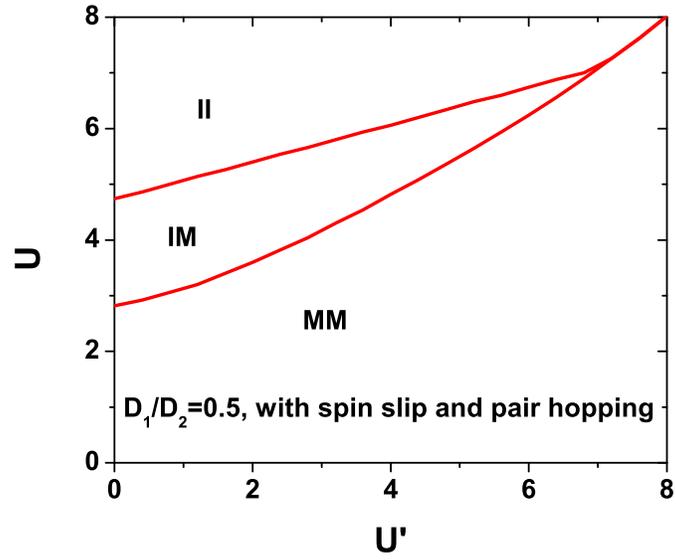}
\end{center}
\caption{The phase diagram for the 2-band Hubbard model with the full Hund's
rule coupling. }
\label{fig3}
\end{figure}

We found that the OSMP appears only when the bandwidth ratio is below a
certain critical value, which is found to be around $D_{2}/D_{1}<0.25$. This
result is in good agreement with that of the slave spin mean field theory%
\cite{Georges}. The role of the Hund's rule coupling has also been
discussed by many groups within the frame of DMFT\cite{Koga,
Liebsch}. With the increment of the Hund's rule coupling, the area
of the OSMP in the phase diagram get larger and larger. In Figure
\ref{fig3} we plot the phase diagram with the full Hund's rule
coupling including both the spin flip and pair hopping terms in
equation \ref{3band} .

\begin{figure}[tbp]
\begin{center}
\includegraphics[width=10cm,angle=0,clip=]{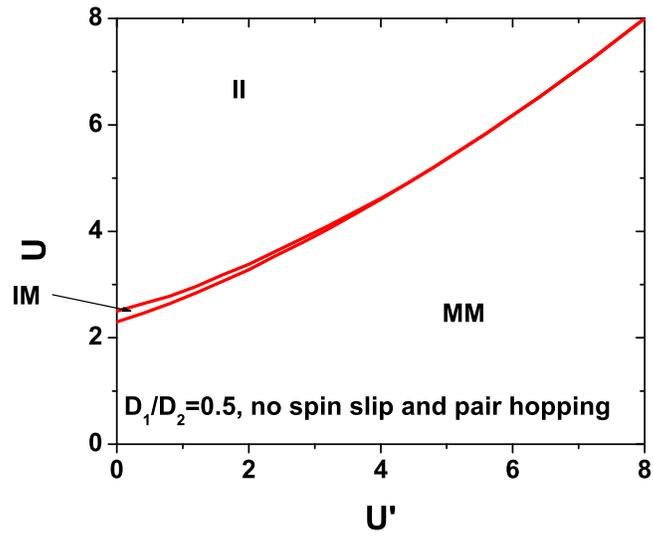}
\end{center}
\caption{The phase diagram for the 2-band Hubbard model with only
longitudinal terms in Hund's rule coupling.}
\label{fig4}
\end{figure}

The transverse part of the Hund's rule coupling including the spin-flip and
pairing hopping terms plays an extremely important role in OSMT. The OSMP is
strongly suppressed without these two transverse terms. This important
feature can be also captured by the present slave boson mean field theory.
As shown in Figure \ref{fig4}, the area of OSMP is shrunk dramatically when
the two transverse terms are switched off, which is also in good agreement
with the DMFT results.

\bigskip

\section{OSMT in the three band Hubbard model}

\bigskip

\begin{figure}[tbp]
\begin{center}
\includegraphics[width=10cm,angle=0,clip=]{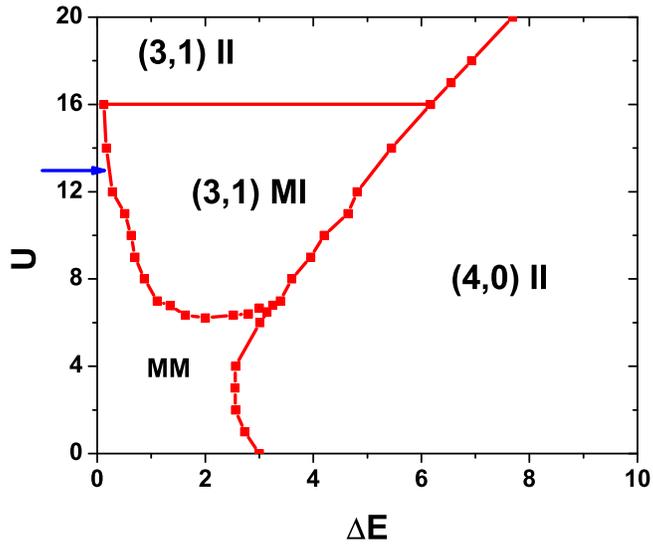}
\end{center}
\caption{ The phase diagram for the OSMT in the 3-band Hubbard
model with full Hund's rule coupling $J=0.25U$. The x-axis is the
crystal field splitting and y-axis is the intra band local
interaction $U$.} \label{fig5}
\end{figure}

\begin{figure}[tbp]
\begin{center}
\includegraphics[width=10cm,angle=0,clip=]{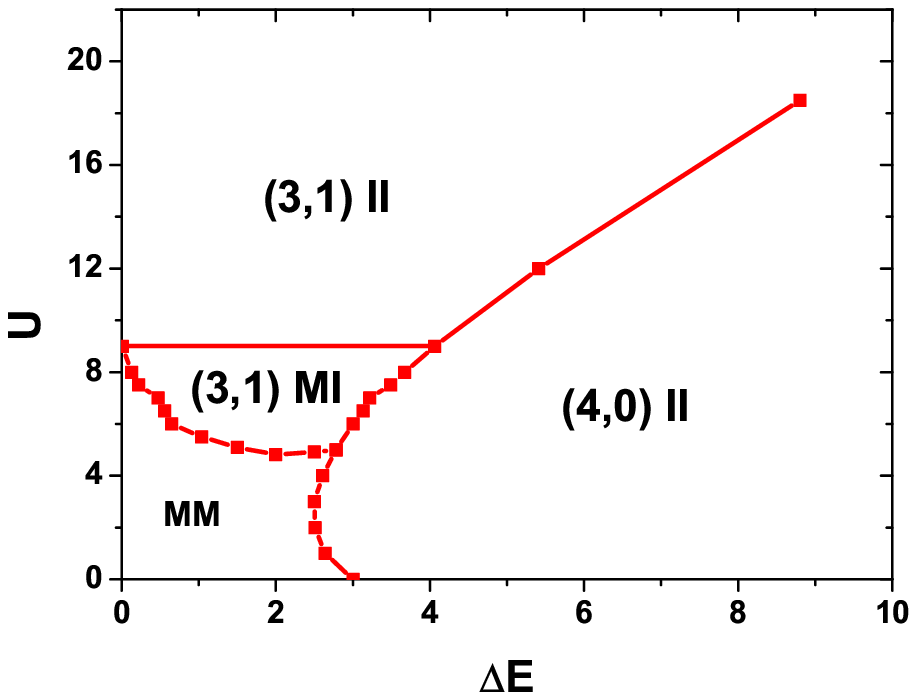}
\end{center}
\caption{The phase diagram for the OSMT in the 3-band Hubbard
model with only longitudinal Hund's rule coupling $J=0.25U$. The
x-axis is the crystal field splitting and y-axis is the intra band
local interaction $U$.} \label{fig6}
\end{figure}

\begin{figure}[tbp]
\begin{center}
\includegraphics[width=10cm,angle=0,clip=]{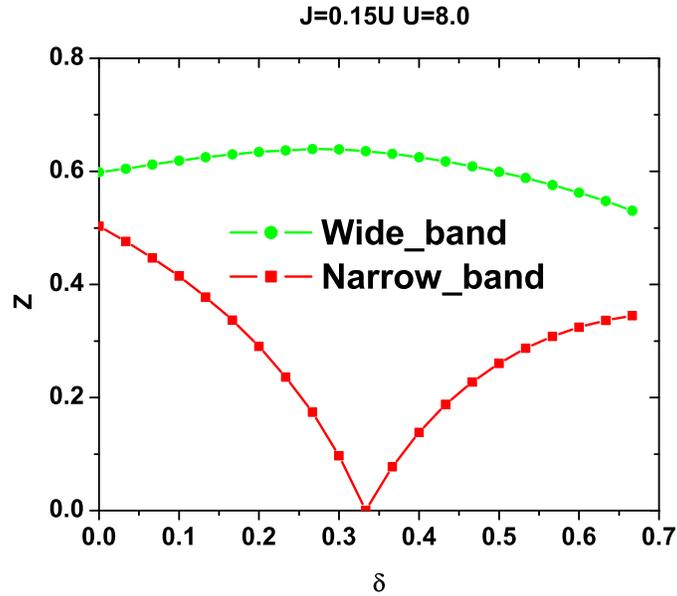}
\end{center}
\caption{The quasi-particle weight for the wide band (circles) and narrow
band (squares) as the function of $\protect\delta$, where $\protect\delta%
=n_{narrow}-{\frac{4}{3}}$. }
\label{fig7}
\end{figure}

The theoretical study of the orbital selective Mott transition is motivated
by the surprising Curie-Weiss behavior of the spin susceptibility in
CaSrRuO4, which is unexpected because the transport measurement shows
metallic behavior. The two band Hubbard model with Hund's rule coupling is
proposed as the simplest toy model for OSMT. While compared to the realistic
situation in CaSrRuO4, which has four electrons occupying three t2g
orbitals, the two band model is over simplified and can not capture a very
important feature of OSMT in $Sr_{x}Ca_{2-x}RuO_{4}$. Unlike the situation
in the 2 band model, the OSMT in $Sr_{x}Ca_{2-x}RuO_{4}$ is accompanied by
the charge redistribution even without crystal field splitting. Since the
average occupation number in $Sr_{x}Ca_{2-x}RuO_{4}$ is 4/3 per orbital per
site, which is away from the half filling, 1/3 of electron has to be moved
from the wide bands to the narrow band to make it half filled and
insulating. Therefore the crystal field splitting, which always reduces the
tendency of OSMT in the two band model, can greatly enhance the OSMT in the
three band model, because it can induce such charge redistribution required
by OSMT. In the present paper, we have systematically studied the three-band
Hubbard model with Hund's rule coupling using the slave boson mean field
approach. Since the optical conductivity measurement\cite{exp1} indicate the
local moment is on the $d_{xy}$ orbital, we choose the band width of the
three different bands in eq.\ref{3band} to be $2D,2D$ and $D$. The phase
diagram for the full Hund's rule coupling is plotted in figure \ref{fig5}
with the interaction strength U and crystal field splitting $E=\Delta
_{1}-\Delta _{3}$ being the y and x axis respectively. The $(3,1)II$ phase
represents the Mott insulator phase for both wide and narrow bands with
occupation numbers of wide band and narrow band to be 3 and 1 respectively. $%
(3,1)MI$ phase represent the metallic phase for the wide band and
insulating phase for the narrow band. $MM$ phase represents the
metallic phase for all the three bands. And $(4,0)II$ phase
represents the band insulator phase with all the four electrons
filling the two wide bands. We find that the OSMT never occurs for
this model without crystal field splitting. Small crystal field
splitting will induce an OSMT very effectively indicating that the
metallic state is unstable against the crystal field in this
regime. Since the crystal field splitting is induced by lattice
distortion, our result indicates that the OSMT in
$Sr_{x}Ca_{2-x}RuO_{4}$ is not purely interaction driven. Instead,
the lattice degree of freedom will also paly a very significant
role. To further verify this point, we plot the quasi-particle
weight as the function of the amount of charge transfer among
$t_{2g}$ orbits, which is determined by the crystal field
splitting, in figure\ref{fig7}. We can see very clearly that the
quasi-particle weight goes to zero when the charge transfer
$\delta $ reaches $1/3$, which indicates a new type of Mott
transition driven by the crystal field splitting. When switching
off the spin flip and pair hopping terms, the over all feature of
the phase diagram keeps unchanged, especially the OSMT does not
occur without the crystal field splitting. The area of OSMP in the
phase diagram decreases dramatically as shown in figure \ref{fig6}
which is quite similar to the two band model.

\bigskip

\section{Concluding remarks}

The OSMT in three-band Hubbard model with different band width is
studies in detail using the slave Boson mean field theory. First
we generalize the slave Boson approach proposed by G. Kotliar and
A. E. Ruchenstein\cite{gabi} to the multi-band system with full
Hund's rule coupling terms and benchmark it with the other
numerical approaches like DMFT in the two band Hubbard model. Then
we apply it to study the possible OSMT in the three-band Hubbard
model with both Hund's rule coupling terms and crystal field
splitting. Our numerical result shows that unlike the case in the
two-band model the crystal field splitting plays a very important
role in stabilizing the OSMP. In fact a new type of OSMT, which is
driven by the crystal field is observed in our study. Since the
crystal field splitting is induced by the lattice distortion, our
results strongly suggest that the lattice degree of freedom
is plays a crucial role in the OSMT in the compound like $%
Sr_{x}Ca_{2-x}RuO_{4}$.

\end{document}